# Enduring Access to Rich Media Content: Understanding Use and Usability Requirements

Madeleine Casad, Oya Y. Rieger and Desiree Alexander
Cornell University Library
{mir9, oyr1, dca58}@cornell.edu

## Abstract

Through an NEH-funded initiative, Cornell University Library is creating a technical, curatorial, and managerial framework for preserving access to complex born-digital new media objects. The Library's Rose Goldsen Archive of New Media Art provides the testbed for this project. This collection of complex interactive born-digital artworks are used by students, faculty, and artists from various disciplines. Interactive digital assets are far more complex to preserve and manage than single uniform digital media files. The preservation model developed will apply not merely to new media artworks, but to other rich digital media environments. This article describes the project's findings and discoveries, focusing on a user survey conducted with the aim of creating user profiles and use cases for born-digital assets like those in the testbed collection. The project's ultimate goal is to create a preservation and access practice grounded in thorough and practical understanding of the characteristics of digital objects and their access requirements, seen from the perspectives of collection curators and users alike. We discuss how the survey findings informed the development of an artist questionnaire to support creation of user-centric and cost-efficient preservation strategies. Although this project focuses on new media art, our methodologies and findings will inform other kinds of complex born-digital collections.

## 1 Introduction

Despite its "new" label, new media art has a rich 40-year history, making obsolescence and loss of cultural history an imminent risk. As a range of new media are integrated in art works, these creative objects are becoming increasingly complex and vulnerable due to dependence on many technical and contextual factors (Delve, *et al*., 2012). The phrase "New media art" denotes a range of creative works that are influenced or enabled by technological affordances. The term also signifies a departure from traditional visual arts (e.g., paintings, drawings, sculpture, etc.). Another characteristic of new media art that adds further complications to the preservation process is its interactive nature. Works in this genre often entail, and indeed rely on, interactions between artists and viewers/observers.

In 2013, Cornell University Library received a research and development grant from the National Endowment for the Humanities to design a framework for preserving access to digital art objects. The Preservation and Access Frameworks for Digital Art Objects (PAFDAO) was undertaken in collaboration with Cornell University's [Society for the Humanities](#) and the [Rose Goldsen Archive of New Media Art](#), a collection of media artworks housed in the Library's Division of Rare and Manuscript Collections. The project aims to develop scalable technical frameworks and associated tools to facilitate enduring access to complex, born-digital media objects, working primarily with a test bed of nearly 100 optical discs from the holdings of the Goldsen Archive. The preservation model developed will apply not merely to new media artworks, but to other rich digital media environments (for instance see Kirschenbaum, *et al.*, [2010](#)). Many of the issues we have been addressing within the framework of this project apply to other rich digital contents, not limited to artistic productions.

From the beginning, the project team has recognized that both metadata frameworks and access strategies would need to address the needs of future as well as current media art researchers. Toward that end, we developed a survey targeting researchers, artists, and curators to expand our understanding of users and use cases. This article summarizes key findings of the survey and describes their impact on our current preservation and access frameworks and future plans.[1]

## 2 About the Collection

The ultimate aim of the PAFDAO project is to create generalizable new media preservation and access practices that will be applicable for different media environments and institutional types. The nature of the project's test collection, a set of CD-ROM artworks from Cornell's Rose Goldsen Archive of New Media Art[2], has meant that the project provides a case study in new media preservation that may be informative to library and museum contexts alike.

Rose Kohn Goldsen (1917-1985) was a professor of Sociology at Cornell University and an early critic of commercial mass media's impact on social and ethical imagination. Named in her honor, the Rose Goldsen Archive of New Media Art was founded in 2002 by Professor Timothy Murray (Director, Society for the Humanities, Cornell University) in the Cornell Library Division of Rare and Manuscript Collections as an international research collection for scholars of new media and media art history (Murray, [2008](#)). Since its founding, the Goldsen Archive has grown to achieve global recognition as a prominent research collection that documents more than 60 years of the history of aesthetic experimentation with electronic communications media. These collections span the two most crucial decades in the emergence of digital media art, from 1991 to the present, tracing the historical shift in emphasis within media culture from disc-based to networked and Web-based applications. They also mark the early stirrings of a networked, interactive digital culture that has subsequently become the global norm. The Goldsen Archive constitutes a vital record of our cultural and aesthetic history as a *digital* society.

The PAFDAO project focused on a subset of born-digital media artworks on CD-ROM. These artworks were created for small-screen, single-user experience, and dated back as far as the early 1990s. The cultural significance of such artworks is great. Among other things, they represent the early development of interactive interfaces that are now a major part of our everyday life. And artists' exploration of the expressive possibilities these new multimedia interfaces have to offer. Despite their cultural value, and their relatively recent production, such artifacts present serious preservation challenges and obsolescence risks.

To begin with, no archival best practices yet exist for preserving such assets. Many are stored on fragile storage media like optical discs, meaning that physical damage as well as data degradation or "bit rot" pose serious dangers to the integrity of the information. In the case of the PAFDAO project's test collection, many of these discs were artist-produced and irreplaceable.

Interactive digital assets are, furthermore, far more complex to preserve and manage than single, uniform digital media files. A single interactive work can comprise an entire range of digital objects and dependencies, including media files in different types and formats, applications to coordinate the files, and operating systems to run the applications. If any part of this complex system fails, the entire asset can become unreadable. This danger is especially acute in the case of artworks. In most cases, interactive digital artworks are designed to create unique, multimedia experiences for users. An even relatively minor problem with an artwork's rendering—for example, an obsolete media player that no longer operates as expected—has the potential to significantly compromise an artwork's "meaning." Simply migrating information files to another storage medium is not enough to preserve their most important cultural content. When the PAFDAO project began, approximately 70 percent of the artworks in the test collection could not be accessed at all without using legacy hardware—a specialized computer terminal that runs obsolete software and operating systems.

The project's objective was to provide "best-feasible" access to artworks, and document the distance between "feasible" and "ideal," as well as we could understand it. Very soon after beginning PAFDAO, the project team realized that, contrary to our initial assumptions, operating system emulation would be a viable access strategy at scale for our complex digital media holdings (for information about emulation, see Lange, 2012). Embracing emulation as an access strategy meant that the team could provide better access more easily to more artworks in the collection. Though increasingly feasible, however, emulation is not always an ideal access strategy: emulation platforms can introduce rendering problems of their own, and emulation usually means that users will experience technologically out-of-date artworks with up-to-date hardware. This made it all the more important for the team to survey media art researchers, curators, and artists, in order to gain a better sense of the relative importance of the artworks' most important characteristics for different kinds of media archives patrons.

## 3 About the Survey

We developed a questionnaire that presented users of media archives with a number of open-ended, largely qualitative and non-restrictive questions about their needs, goals, and preferences. In January 2014, we circulated the questionnaire on several preservation, art, and digital humanities mailing lists.

The PAFDAO team initially hoped that survey results would support the identification of "personas," or broad profiles of media archives users who shared similar needs and preferences. We hoped that these profiles would direct both metadata framework and access provisions. As it happened, no such clear classifications emerged, yet questionnaire results were still vastly informative, and shaped the development of the PAFDAO project in integral ways. In the remainder of this paper, we offer an overview of noteworthy trends and comments, then discuss the conclusions we draw from these results and their impact on the PAFDAO workplan and preservation framework.

## 4 Survey Results

A total of 170 people responded to the questionnaire. Respondents came from disparate geographical locations, including the US, Germany, France, UK, Australia, and Argentina. Of 170 respondents, 122 responded as an individual researcher or practitioner and 48 responded on behalf of an archive, museum, or other cultural heritage institution.[3] We did not observe any significant differences in the responses of these two groups (personal and institutional responses), possibly due to the fact that even at an institutional level, new media projects and collections are led by small, specialized teams of committed individuals. Respondents often held multiple roles and characterized themselves non-exclusively as artists (48%), researchers (47%), educators (25%), curators (20%), collection specialists (24%). The scope of digital media art collections respondents worked with was also broad, and included digital installation, video and images, interactive multimedia, audio, 3-D visualization, and websites.

The key impetus behind the survey was to understand what kind of research questions and needs were motivating users to search for and use media works. This information is critical for the research team to identify and assess the nature and extent of viewing experience that needs to be preserved. In aggregate, respondents gave almost equal weight to artistic, social, historical, cultural, aesthetic, and technical research frameworks. Several described pedagogical uses and how they use media works in teaching and learning. Some sample research questions include:

- How are technologies assisting the exploration of political issues by artists?
- How do you bring the work to the viewer through the interactive power of technologies?
- Do digital works explore something further than the analog approaches can do?
- How do technologies support and stimulate community engagement?
- How are access issues for individuals with lower economic backgrounds being addressed?

- What are the possible implications of gender in digital media artworks?
- What does it mean to view an art work that is designed for an old TV set in a larger installation?

The respondents cited a number of serious impediments they had encountered in conducting research involving new media art. For example, they mentioned the lack or insufficiency of documentation and metadata, discovery and access provisions, and technical support. Ones who use new media collections in support of teaching and learning listed several impediments such as vanishing webpages, link rot, poor indexing, gap for works from the 80s and 90s, and the lack of quality documentation. Also often underscored were the complexity of legal issues and access rights. One respondent pointed out that, due to a widespread "disinterest in preserving the cultural artifacts of the digital age," there is a lack of understanding of the importance of these objects for cultural history. Another comment noted infrequent access requests and therefore difficulties in justifying institutional investment in preservation efforts for future use.

One of the respondents wrote, "In a society that is rushing headlong into the future, it is vital that we preserve the efforts of those who have early works in this new culture." Another one commented that as technologies evolve, some works become very easy to create and therefore some users don't understand the significance of a work and how it was a complicated piece to produce at the time. Such sentiments underscore the importance of documenting cultural context to situate the work from artistic, historic, and technical perspectives.

For practicing artists, there were several concerns about the longevity of their creative work. Some expressed concern about the difficulty of selling works that may become obsolete within a year. Many worried that it was difficult to store or archive immersive installations, interactive pieces, and work with dependency on external files. They also mentioned copyright issues as a significant challenge. Many emphasized the importance of historical contexts, usability, and discovery. One of them pointed out that archiving has become a part of his practice and he feels the pressure to consider future uses as he is going through a creative process.

For curators of new media art, many indicated that they don't include born-digital interactive media in their holdings because either such materials fall outside of collecting scope or the procedures for providing access are too complex or unsustainable. For those who collect this genre, the biggest concerns were trying to identify which aspects of interaction experiences to preserve and how to capture as much information as possible to assist future users. Out of the twenty survey respondents who answered on behalf of an educational or cultural institution, only one organization could claim a sophisticated and integrated web-based discovery, access, and preservation framework. The others indicated that access needed to be arranged through a special arrangement such as setting an appointment. They cited a range of preservation strategies they rely on, including migration, creation of search and discovery metadata, maintaining a media preservation lab, providing climate controlled storage, and collecting documentation from the artists.

## 5 Content Authenticity and Authentic User Experience

As mentioned above, the PAFDO survey of users of media archives did not, as we had hoped, result in the definition of clear user profiles or personas. However it had several important effects on the PAFDAO project. First, we noted a significant concern among our respondents for "authenticity"—understood as a cultural rather than technical concept.

The International Research on Permanent Authentic Records in Electronic Systems (InterPARES) project defines an authentic record as "a record that is what it purports to be and is free from tampering or corruption" (MacNeil, *et al*., 2001, referenced in Dietrich & Adelstein, 2015). Verifying the bit-level self-identity of a digital object over time can be accomplished relatively easily with checksums, automated fixity checks, and collection audits. When working with cultural artifacts, however, "authenticity" becomes a more nebulous and controversial concept. Conservation measures undertaken to restore an artwork to some approximation of its original appearance may, in fact, alter its original form in ways that can affect its meaning. This is especially true in the case of artworks conceived to be ephemeral or experiential, or works that involve "contemporary" technologies that become obsolete, even obscure, over time.

Our questionnaire respondents seemed to respect this difficulty. Reading across the complete pool of responses, we noted that the desired sense of "authenticity" derived not from some naïve sense of the object's pristine originality, but rather from a sense that the archiving institution has made a good-faith commitment to ensuring that the artist's creative vision has been respected, and providing necessary context of interpretation for understanding that vision—and any unavoidable deviations from it.

We had excellent models for addressing these concerns. Within the last ten to fifteen years, many arts organizations have joined forces to develop shared practices for the conservation of technology-based media, but also difficult-to-document arts such as performance, video art and multi-media installations. Examples include Independent Media Arts Preservation (IMAP); The Variable Media Network; Matters in Media Art (a collaborative project between the Tate, the New Art Trust (NAT) and its partner museums—the Museum of Modern Art (MoMA), the San Francisco Museum of Modern Art (SFMOMA)); and INCCA (International Network for the Conservation of Contemporary Art).[4] The most significant commonality of these initiatives is their shared emphasis on appropriate documentation. While some complex time-based artworks can never be authentically replicated, it is generally agreed that, with proper documentation, many can be reinterpreted, adapted and revived for modern audiences. In cultural heritage organizations, this documentation can take the form of technical and descriptive metadata tailored for the breadth and specificity of new media, detailed installation instructions, detailed exhibition histories, and so forth. Above all, practices for working directly with artists have been especially important conservation tools, and the initiatives cited above provide excellent models for how artist interviews can aid efforts to preserve complex artworks; see, for example, the Variable Media Questionnaire (Depocas, *et al*., 2003).

In response to these considerations raised by our user survey, we developed a conservation-oriented artist questionnaire and interview process, pushing the integration of archival protocols as far upstream as possible, to the point of content creation and initial curation. Enlisting the help of our project advisors, we worked with existing models, but adapted these models significantly. We streamlined and simplified our artist questionnaire to address specific aspects of our emerging preservation and access framework. We were particularly concerned about communicating with artists and enlisting their input about our decision to rely on operating system emulation as a default access strategy. Though easy and readily scalable, emulation introduces variations into the rendering of artworks that artists might not have anticipated; it was clear that we would need to work with artists wherever possible to ensure that artworks' most significant properties and interpretive contexts were preserved, and not obscured, by our access measures.

## 6 Artist Questionnaire

The PAFDAO questionnaire is designed to be a first step in a two-part process, gathering essential information but also laying the groundwork for a more conversational interview process where possible.

First and foremost, the questionnaire elicits artists' input in identifying the most significant properties of individual media artworks by asking about the artists' initial vision for the work, and by posing open-ended questions about the relationship between artistic vision, technology, and historical contexts.

The questionnaire also asks fundamental technological questions. (e.g., "What software or programming language was used to create this artwork?" "What hardware and software were optimal for running this artwork when it was new?") We inquire as to whether artists still have the working files they used in creating the artwork, including source code; these would constitute a deep technological and historical context for the works, and also an invaluable resource for future conservation work (Engel & Wharton, 2014). We also ask about related artworks or websites, and whether any of these materials may have been archived by another person or institution. Networks of collaboration between archiving institutions will become more and more important in preserving cultural, historical, and technological contexts of reference that will be essential to understanding these artworks.

The questionnaire also discloses foreseeable problems in our chosen access frameworks, including specific rendering issues that might come about with different emulation platforms:

*We have found virtual machine emulation to be an effective strategy for providing research access to interactive digital artworks. Running older artworks in an emulation environment*

*may involve changes to the look and feel of the original artwork. Our default access strategy is likely to involve:*

- *Current, commercial-grade hardware and peripherals (mouse, screen, keyboard, etc.)*
- *Color shift associated with the change from CRT to LED monitor screens*
- *Possible alterations to the speed of animation and interactive responsiveness*
- *Possible changes to audio quality*
- *Presentation of digital surrogates rather than original physical materials that may have accompanied the artwork (discs, booklets, cases, etc.)*

We ask artists to describe how such changes might affect their initial vision for the work. We also request permission to provide works in emulation, outline the kinds of documentation we expect to provide archive users, and invite artists to work with us on supplementary or alternate forms of documentation if they choose:

*We expect to present users with a general statement about the effects of our emulation environments on the rendering of an artwork.*

*If you would like to author or co-author a more specific statement about how these changes may affect your work, we can provide researchers with this information as well. In some cases, we may be able to provide additional documentation of original rendering conditions. Please let us know if you would like to discuss these possibilities further.*

Finally, the questionnaire furthermore provides us with an opportunity to revisit rights agreements, which must be updated in light of new access technologies, and an opportunity to invite further conversation (a follow-up interview) and collaboration with the artist.

## 7 Concluding Remarks

A reoccurring theme in our findings involved the difficulties associated with capturing sufficient information about a digital art object to enable an authentic user experience. This challenge cannot and should not be reduced to the goal of ensuring bit-level fixity checks or even providing technically accurate renderings of an artwork's contents as understood on the level of individual files. As Rinehart & Ippolito ([2014](#)) argue, the key to digital media preservation is variability, not fixity. The trick is finding ways to capture the experience—or a modest proxy of it—so that future generations will get a glimpse of how early digital artworks were created, experienced, and interpreted. So much of new media works' cultural meaning derives from users' spontaneous and contextual interactions with the art objects. Espenschied, *et al*. ([2013](#)) point out that digital artworks relay digital culture and "history is comprehended as the understanding of how and in which contexts a certain artifact was created and manipulated and how it affected its users and surrounding objects." For a work to be understood and appreciated, it is essential for the archiving institution to communicate a cultural and technological framework for interpretation. As one user survey respondent noted, some works that come across as mundane now may have been among the highly innovative trailblazers of yesterday. Given the speed of technological advances, it will be essential to capture these historical moments to help future users understand and appreciate such creative works.

The PAFDAO survey of users of media archives affirmed the importance of institutions like the Rose Goldsen Archive, which is able to provide a breadth of media technological, historical, and cultural contexts to researchers and educators through its extensive and accessible collections.[5] It also underscored the need for archiving institutions to be in contact with one another, and to be conscious of the need for greater integration of discovery and access frameworks across multiple institutions as they move forward in developing new preservation plans and access strategies for their collections. Providing appropriate cultural and historical contexts for understanding and interpreting new media art is part of each institution's individual mission, but also a matter of collective importance, given the rarity of such collections, the numerous challenges of establishing preservation protocols, and the overall scarcity of resources. As we conclude, we must emphasize that, as artists have increasing access to ubiquitous tools and methodologies for creating complex art exhibits and objects, we should expect to see an increasing flow of such creative works to archives, museums, and libraries. It is nearly impossible to preserve these works through generations of technology and context changes. Therefore, diligent curation practices are going to be more essential than ever in order to identify unique or exemplary works, project future use scenarios, assess obsolesce and loss risks, and implement cost-efficient strategies.

## Acknowledgements

We would like to express our gratitude to the National Endowment for the Humanities for supporting this project, to the project advisory board, to consultants Chris Lacinak, Kara VanMalssen, and Alex Duryee of AVPreserve, and to the PAFDAO project team, including Timothy Murray (co-PI), Dianne Dietrich, Desiree Alexander, Jason Kovari, Danielle Mericle, Liz Muller, Michelle Paolillo.

## Notes

[1] An early version of this report is available at DSPS Press: the blog of Cornell University Library's Division of Digital Scholarship and Preservation Services. See "Interactive Digital Media Art Survey: Key Findings and Observations: DSPS Press".

[2] The Goldsen Archive's holdings range to include media formats such as reel-to-reel videotape, floppy disk, database artworks housed on external hard drives, and works of

net.art. All of these formats pose unique and significant preservation challenges. For more information, please see the [Goldsen Archive](#) website.

[3] Out of 170 respondents, 80 fully and 32 partially completed the survey, and 58 took a quick look without responding. We suspect that the incomplete survey indicates a combination of curiosity and unfamiliarity with the program area, as media art research, curation, and practice still constitute fairly specialized fields. Only twenty-four respondents indicated that their institutions include born-digital interactive media artworks and artifacts in their holdings. Several respondents who identified as curators indicated that born-digital interactive media would fall outside the scope of their collections. In some cases, they also noted that procedures for providing access to such materials are prohibitively complex or unsustainable.

[4] For further information and documentation please see http://imappreserve.org/, http://variablemedia.net/, http://www.tate.org.uk/about/projects/matters-media-art, and http://www.incca.org/

[5] Cornell University Library's commitment to provide broad and democratic access to its special collections was a key reason why founding Goldsen Archive curator Timothy Murray located the Goldsen collections within the library. Cornell's Division of Rare and Manuscript Collections has notably open policies for user access; see http://rmc.library.cornell.edu/ for more information.

## References

[1] Delve, J. *et al*. (2012). The Preservation of Complex Objects. Volume One: Visualizations and Simulations.

[2] Depocas, A., Ippolito, J., Jones, C., eds. (2003). Permanence Through Change: The Variable Media Approach. Guggenheim Museum Publications, NY & Daniel Langlois Foundation, Montreal.

[3] Dietrich, D., Adelstein, F. Archival science, digital forensics, and new media art. Volume 14, Supplement 1, August 2015, Proceedings of the Fifteenth Annual DFRWS Conference. http://doi.org/10.1016/j.diin.2015.05.004

[4] Espenschied, D., Rechert, K., Valizada, I., Von Suchodoletz, D., Russler, N. (2013). "*Large-Scale Curation and Presentation of CD-ROM Art*", iPRES 2013.

[5] Engel, Deena, and Glenn Wharton (2014). Reading between the Lines: Source Code Documentation as a Conservation Strategy for Software-Based Art. Studies in Conservation 59(6): 404—415. http://doi.org/10.1179/2047058413Y.0000000115

[6] Kirschenbaum, M. et al. (2010). Digital Forensics and Born-Digital Content in Cultural Heritage Collections. CLIR.

[7] Lange, A. (2012). KEEP Strategy Paper.

[8] MacNeil, H. Wei, C., Duranti, L., Authenticity Task Force Report. InterPARES.

[9] Murray, T. (2008). Thinking Electronic Art Via Cornell's Goldsen Archive of New Media Art. NeMe: The Archival Event.

[10] Rinehart, Richard, and Jon Ippolito, (2014) Re-Collection: Art, New Media, and Social Memory. Leonardo. Cambridge, Massachusetts: The MIT Press.

## About the Authors

**Madeleine Casad** is Curator for Digital Scholarship at Cornell University Library. As Associate Curator of the Rose Goldsen Archive of New Media Art, she manages an exciting collection of media objects that present a wide range of preservation and access challenges. She coordinates many of the Library's Digital Humanities initiatives, and plays a leading role in education and outreach programs to promote the innovative use of digital collections in humanities scholarship. She holds a PhD in Comparative Literature from Cornell University.

**Oya Y. Rieger** is Associate University Librarian for Scholarly Resources and Preservation Services at Cornell University Library. She provides leadership for full lifecycle management of scholarly content, including selection, creation, design, maintenance, preservation, and conservation. She is interested in current trends in scholarly communication with a focus on needs assessment, requirements analysis, business modeling, and information policy development. She holds a PhD in Human-Computer Interaction (HCI) from Cornell University.

**Desiree Alexander** is the PAFDAO Collections Analysis Assistant and has worked with the Goldsen Archive since 2012, assisting with the Goldsen's experimental video and digital media preservation projects. She is also co-lead in surveying Cornell's A/V assets to locate at risk materials campus-wide in an effort to develop preservation and access strategies. She holds a MS in Information Studies and an MA in Public History from SUNY Albany, and an undergraduate degree in Art History from Ithaca College.